# Molecular Dynamics Study of Irradiation-Induced Defect and Dislocation Evolution in Strained Nickel

Maciej Wilczynski
Mark Fedorov
Tymofii Khvan
F. Javier Dominguez-Gutierrez
Jacek Jagielski

**Abstract**

Molecular dynamics simulations were performed to quantify how elastic tensile strain alters irradiation damage in fcc Ni single crystals subjected to cumulative overlapping displacement cascades initiated by 5 keV PKAs at 300 K. Each cascade produces (54) Frenkel pairs on average, in close agreement with the NRT estimate, and a sequence of ten cascades corresponds to low irradiation dose in the present cell. Comparing representative cascades (1, 5, and 10) shows that all ballistic and heat–spike recovery processes complete within (≲10) ps, validating a 20 ps cascade window. Tensile strain systematically increases retained defect populations relative to the unstrained case, consistent with elastic stabilization of interstitial-rich configurations. The heat-spike duration and peak disorder decrease monotonically with strain, indicating faster local energy dissipation under stress. Dislocation analysis reveals a strong bias toward Shockley partial loops; Hirth, stair–rod, Frank, and perfect components remain secondary. Total and Shockley dislocation densities converge to a steady–state saturation of ($\sim 10^{16}, \mathrm{m}^{-2}$), and their strain dependence is well captured by a Kocks-–Mecking description with a saturation density of ($\rho_{ss} \approx 1 \times 10^{-2}$) in normalized units. These results demonstrate robust strain-–irradiation coupling in fcc metals at low doses.

Maciej Wilczynski Mark Fedorov Tymofii Khvan F. Javier Dominguez-Gutierrez Jacek Jagielski
NOMATEN Centre of Excellence, National Centre for Nuclear Research, ul. A. Soltana 7 05-400 Otwock, Poland
Email Address: javier.dominguez@ncbj.gov.pl

## 1 Introduction

Materials used in nuclear environments operate under harsh conditions that combine neutron irradiation, thermomechanical stresses, and corrosive effects, posing significant challenges to component reliability. Their mechanical properties progressively degrade under sustained neutron fluence due to irradiation-induced microstructural changes (Gaganidze and Aktaa 2013; Slugeň et al. 2020; Wakai et al. 2021). These changes are driven by the formation and evolution of defect structures such as dislocation loops, voids, cavities, precipitates, and, in face-centered cubic (FCC) metals, stacking-fault tetrahedra (Dethloff et al. 2016; Schäublin et al. 2005). The accumulation of such defects leads to phenomena including hardening, embrittlement, void swelling, irradiation creep, and strain softening (Odette and Lucas 1997; Garner et al. 1981; Kohyama et al. 1994; Chaouadi 2008), ultimately reducing the structural integrity of reactor components and increasing the risk of premature failure.

Post-irradiation examinations provide essential information on the degradation of mechanical performance through macroscopic testing and microstructural characterization by advanced microscopy. However, these experiments are costly and time-consuming, requiring long irradiation campaigns, limited irradiation space, and extensive shielding due to induced radioactivity. Alternatively, ion and proton irradiation can mimic neutron damage, offering accelerated dose rates, reduced costs, and the absence of residual activity (Was 2015; Gentils and Cabet 2019). Still, understanding the formation, evolution, and interaction of primary defects relies heavily on computational modeling and theoretical frameworks (Jagielski and Thomé 2009).

Among the available computational approaches, molecular dynamics (MD) simulations have proven to be a powerful tool for studying atomic-scale processes associated with irradiation damage (Azeem et al. 2018; Khater et al. 2014; Gao and Weber 2003; Aligayev et al. 2025; Mieszczynski et al. 2025; Mieszczynski et al. 2024; **Cichocki_Dominguez-Gutierrez_Wyszkowska_Kurpaska_Muszka_2025?**). MD enables the direct observation of displacement cascades, defect clustering, and dislocation formation, offering insight into the transient processes that govern defect accumulation and recovery (Wyszkowska et al. 2025; Ustrzycka et al. 2025). Moreover, MD plays a key role within multiscale modeling strategies, where atomistic results provide constitutive parameters for mesoscale and continuum-scale models (Lin et al. 2022; Kumar et al. 2012; Ustrzycka et al. 2024). This framework underpins several international research efforts dedicated to predicting the performance and lifetime of nuclear structural materials, such as the M4F (Malerba et al. 2021), GETMAT (Fazio et al. 2011), and PERFECT (Massoud et al. 2010) projects. The present study investigates the effect of tensile strain on irradiation-induced defect evolution in nickel by performing molecular dynamics simulations of cumulative overlapping 5 keV collision cascades at 300 K. The work focuses on understanding how applied mechanical strain modifies defect formation, recovery dynamics, and dislocation evolution during successive cascade events. Unlike previous studies, which primarily examined irradiation effects in unstrained systems, this work explicitly couples elastic deformation with the irradiation process to capture the interplay between external stress and defect dynamics. The simulations reveal how stress-assisted defect mobility influences heat spike dissipation, defect survival, and the saturation of dislocation density under irradiation. Furthermore, the evolution of the total dislocation density with strain is analyzed using the Kocks–Mecking model, providing, for the first time, a quantitative link between atomic-scale irradiation mechanisms and strain-dependent microstructural stabilization in metallic systems relevant to nuclear applications.

## 2 Computational Methods

In this work, nickel is chosen as a model face-centered cubic (fcc) system to establish a clear atomistic reference for the coupling between mechanical strain and irradiation-induced defect evolution under well–controlled conditions. Pure Ni offers a well-characterized crystallographic structure, reliable interatomic potentials, and extensive literature on cascade damage, which together provide a robust baseline for interpreting the effect of strain on defect production, heat spike dynamics, and dislocation evolution (**Cichocki_Dominguez-Gutierrez_Wyszkowska_Kurpaska_Muszka_2025?**; Kurpaska et al. 2022; Wyszkowska et al. 2025; Levo et al. 2017). The insights obtained here are intended as a first step toward more complex and application-oriented systems, such as Ni-based alloys and austenitic steels, where chemical disorder, solute segregation, and irradiation-induced phase stability will further modify the strain–defect interaction (Ustrzycka et al. 2024, 2025). The present study thus serves as a benchmark for extending strain-assisted irradiation modeling to multicomponent alloys relevant for nuclear structural applications.

The atomistic behavior of Ni samples was simulated using the interatomic potentials based on the modified embedded atom method (MEAM) were employed to depict atom-to-atom interactions (Won-Mi et al. 2018). This task was performed using the Large-scale Atomic Molecular Massively Parallel Simulator (LAMMPS) (Thompson et al. 2022). Interatomic interactions smoothly joined to the Ziegler–-Biersack–-Littmark (ZBL) universal repulsive potential to correctly describe short–range interactions occurring during high-energy collisions. The ZBL potential is applied below an inner cutoff radius of 1.0 Å, corresponding to the regime where electron-cloud overlap makes the interaction purely repulsive. For distances larger than 1.0 Å, the potential is gradually blended into the equilibrium MEAM form using a cubic spline interpolation over the range 1.0–1.5 Å. This spline is constructed to enforce continuity of the potential and its first two derivatives, ensuring smooth forces throughout the transition region. The spline parameters were optimized to reproduce reference formation energies of vacancies and self-interstitial configurations, as well as elastic constants, thereby guaranteeing physical consistency in the defect formation regime. Beyond 1.5 Å, the potential is fully described by the original MEAM functional form with an outer cutoff of 6 Å. This ZBL–EAM joining procedure follows well–established approaches used in cascade simulations, and the resulting potential has been verified to produce stable defect structures and defect energetics consistent with reported experimental and computational values (Bonny et al. 2011; Aligayev et al. 2025).

A pure face-centered cubic (fcc) Ni crystal oriented along the [100] direction was generated with a lattice constant of 0.352 nm. The simulation cell was constructed by replicating the unit cell $40 \times 41 \times 42$ times along the $x$, $y$, and $z$ directions, resulting in dimensions of approximately $14.08 \times 14.43 \times 14.78$ nm and containing over 250,000 atoms. The atomic configurations were relaxed using the Fast Inertial Relaxation Engine (FIRE) minimization algorithm (Guenole et al. 2020) until the maximum atomic force on any atom was reduced below $1 \times 10^{-3}$ meV/Å. This procedure ensured that each sample reached its nearest local energy minimum, representing a mechanically stable configuration prior to the initiation of cascade simulations. During the relaxation process, the convergence criterion for the total energy was set such that the relative change in energy between consecutive minimization steps and the magnitude of the total energy both remained below $10^{-7}$ eV. Furthermore, the norm of the global force vector acting on all atoms was constrained to be less than or equal to $(1 \times 10^{-8})$ eV/Å, guaranteeing precise force convergence and structural equilibrium.

In MD simulations of irradiation damage, neutrons cannot be represented explicitly because their interaction with matter occurs through nuclear scattering rather than interatomic forces. Instead, radiation damage is modeled by imparting a prescribed kinetic energy to a selected lattice atom, generating a primary knock-on atom (PKA). This approach reproduces the fundamental physical event underlying neutron and ion irradiation: an atomic recoil that initiates a displacement cascade. The cascade begins with a few–hundred femtosecond ballistic phase in which the PKA transfers momentum to neighboring atoms, followed by the formation of a localized thermal–spike or heat spike, where a nanometer– scale region reaches near-molten disorder before rapidly cooling and recrystallizing. A PKA energy of 5 keV was selected for this study because it represents the lower end of the recoil–energy spectrum generated in Ni by fission neutrons, while also corresponding to the near–surface recoil energies associated with light-ion irradiation commonly used as an experimental surrogate for neutron damage. At this energy scale, the cascade radius is on the order of a few nanometers, consistent with theoretical scaling between cascade energy and melt–zone size. This ensures that both the ballistic region and the transient heat spike remain fully confined within the simulation cell, preventing interactions with the periodic boundaries. Furthermore, a 5 keV cascade generates a sufficient number of Frenkel pairs to allow clear identification of strain-induced modifications to defect production and recombination dynamics, while avoiding the need for significantly larger simulation boxes required at higher energies.

The simulation cell dimensions were chosen to exceed the maximum displacement excursion of the cascade, and the total simulation time was set to 20 ps, which is sufficient to capture the complete evolution of the displacement cascade. As shown in Figure 1, all cascade phases ballistic displacement, heat–spike formation, recombination, and defect freeze–out—-relax fully within the first 8–10 ps. Displacement cascade simulations were initiated by randomly selecting primary knock-on atoms (PKAs) within the simulation cell, with initial velocities assigned to correspond to a recoil energy of 5 keV. The velocity vector was defined as

$$v_i = (v\sin\theta\cos\phi, v\sin\theta\sin\phi, v\cos\theta),$$

where $(v)$ corresponds to the magnitude of the PKA velocity. The polar and azimuthal angles, ( $\theta$ ) and ( $\phi$ ), were sampled to achieve a uniform distribution over the solid angle. Specifically, the polar angle was generated using the transformation ( $\theta = \arccos(1 - 2\xi_1)$ ), where ( $\xi_1$ ) is a uniform random variable in [0,1], ensuring a probability density proportional to ( $\sin(\theta)$). The azimuthal angle ( $\phi$ ) was sampled uniformly within [0, $2\pi$].

This procedure guarantees a statistically isotropic distribution of PKA directions across the ensemble of simulations (Aligayev et al. 2025). Each recoil event was simulated for 20 ps in an energy-conserving setup, with the cascade core evolving under the NVE ensemble while an outer room temperature NVT shell was applied at the cell periphery to absorb the excess kinetic energy introduced by the PKA. This was followed by an additional 10 ps to allow for stress relaxation and swelling. (Ustrzycka et al. 2024; Levo et al. 2017). Defect formation, including dislocation nucleation, was analyzed using the Dislocation Extraction Algorithm (DXA) for dislocation analysis, as implemented in OVITO (Stukowski 2009). Irradiation-induced defects were modeled through cumulative, overlapping collision cascade simulations generated by 5 keV primary knock-on atom (PKA) recoils at 300 K. Each cascade was allowed to evolve until the associated heat spike which is a transient, highly localized region of elevated temperature resulting from the rapid transfer of kinetic energy from the PKA to surrounding atoms, fully developed and dissipated within the simulation cell. This ensured that the thermal transients decayed to equilibrium before subsequent cascades were introduced, preventing artificial thermal overlap and ensuring realistic defect accumulation dynamics. Periodic boundary conditions were carefully verified to prevent artificial interactions between cascades or across cell boundaries, thereby guaranteeing that the resulting defect structures were intrinsic to the irradiation process rather than artifacts of the simulation setup.

MD simulations were performed under compressive loading with a strain rate of $10^8\ \mathrm{s}^{-1}$ applied along the $z$-direction. A strain rate of $10^8\ \mathrm{s}^{-1}$ was applied along the tensile direction. Although this rate is several orders of magnitude higher than those achievable in experiments (typically $10^3$–$10^4\ \mathrm{s}^{-1}$), such high values are necessary in MD simulations due to the limited nanosecond timescales accessible to atomistic modeling. This approach enables the observation of deformation and defect evolution processes within feasible computational times while preserving the correct qualitative trends associated with strain–dependent behavior. Atomic positions were remapped at each computational step to match the instantaneous dimensions of the simulation cell. The selected strain rate reflects the intrinsic computational constraints of MD simulations, which operate on nanometer-length and nanosecond-timescales. To capture meaningful deformation behavior within these limits, relatively high strain rates are required to induce observable mechanical responses. Although this deviates from experimental strain rates, the resulting qualitative insights remain valuable for elucidating deformation mechanisms at the atomic scale. Displacement-controlled straining was strictly imposed along the loading direction, while a barostat maintained zero stress in the two lateral directions. The axial stress component, $\sigma_{zz}$, was directly obtained from the simulations based on the atomic virial formulation, accounting for the number of atoms and their atomic volumes. The applied uniaxial strain was defined as

$$\varepsilon_z = \frac{L_z - L_{z0}}{L_{z0}},$$

where $L_z$ is the instantaneous cell length along the $z$-axis and $L_{z0}$ is its initial value prior to compression. A constant time step of 2 fs was used throughout all simulations. To address the intrinsic stochasticity of displacement cascades, all overlap cascade simulations were performed with statistically independent initial conditions. The PKA was selected at random atom projectiles, and a sequence of ten overlapping cascades was executed. This ten cascade sequence was then repeated 10 times, each repetition using a newly chosen random projectile atom, resulting in a total of 100 cascades per condition. The defect counts, dislocation densities, and other reported quantities correspond to the average over these independent realizations. Statistical errors were evaluated from the standard deviation across the ten repetitions and were found to be much smaller than the systematic differences induced by the applied strain. The purpose of applying strain in the present work is not to mimic the real temporal evolution of mechanical loading, but to establish a well–defined and mechanically equilibrated deformation field before the onset of each irradiation event. Once the target strain is reached, the system rapidly relaxes its internal stresses through the propagation of elastic waves. In crystalline metals such as Ni, the characteristic elastic relaxation time is on the order of tens of femtoseconds, governed by the speed of sound and the interatomic force constants. This relaxation is significantly faster than the picosecond timescale associated with the ballistic and thermal–spike phases of a collision cascade (Nordlund et al. 2018). Consequently, although the strain is applied at a high rate, the deformation field that interacts with the cascade is effectively static and fully equilibrated. From a physical perspective, plastic deformation in Ni requires either the nucleation of new dislocations or the motion of existing ones. Both mechanisms involve overcoming finite energy barriers associated with the Peierls potential and local stress fields. Under the elastic strains used in this work, these barriers remain sufficiently high that no dislocation activity is triggered prior to irradiation. The collision cascade therefore evolves within a statically strained, but otherwise defect–free, crystal. This approach is consistent with the physical picture of irradiation occurring in a pre–stressed material element, where the externally imposed deformation field modifies defect energetics, recombination probabilities, and defect survival. The influence arises from the magnitude of the imposed elastic strain, not from the rate at which it was applied. Accordingly, the MD procedure employed here provides a physically meaningful framework for investigating the interplay between irradiation damage and externally applied stress.

# 3 Results

In this work, two distinct defect metrics are used to stat with the irradiation induced defects production. The number of defects are considered as the instantaneous population of atoms in non-fcc local environments (HCP, BCC, Other) identified via structural analysis with PTM, and therefore represents the transient cascade disorder during the ballistic and heat–spike stages (Aligayev et al. 2025). This quantity should not be interpreted as the number of Frenkel pairs. Vacancies and self- interstitial atoms (SIAs), i.e., true Frenkel defects, are then quantified separately by Wigner–Seitz analysis after full cascade relaxation, and are reported as surviving point–defect populations. Figure 1 presents the temporal evolution of defect formation during overlapping 5 keV collision cascades in both unstrained and strained Ni single crystals. To improve clarity, only three representative cascades—-cascades 1, 5, and 10—are shown for each condition. For all curves, the defect population exhibits the characteristic stages of a displacement cascade: a sharp initial rise corresponding to the ballistic phase, followed by a rapid decrease associated with recombination during the heat spike and early recovery. In the unstrained system (dashed lines), the defect–time profiles for successive cascades closely overlap. Each cascade reaches a stable plateau within the first 8–10 ps, demonstrating that the ballistic and thermal spike processes are fully relaxed well before 20 ps. This confirms that the accumulation of SIAs and vacancies from earlier cascades does not alter the relaxation behavior of subsequent events and that no residual transient dynamics persist beyond this timescale. The strained system (solid lines) shows higher defect peaks and elevated steady-state levels, reflecting the influence of tensile deformation on defect production and recovery. Nonetheless, the strained cascades also reach a well-defined plateau within the same 8– 10 ps interval. This consistent relaxation behavior across both strained and unstrained conditions validates the choice of a 20 ps simulation window per recoil event: all relevant cascade dynamics— including the heat spike and defect recombination—are fully resolved before the next cascade is initiated.

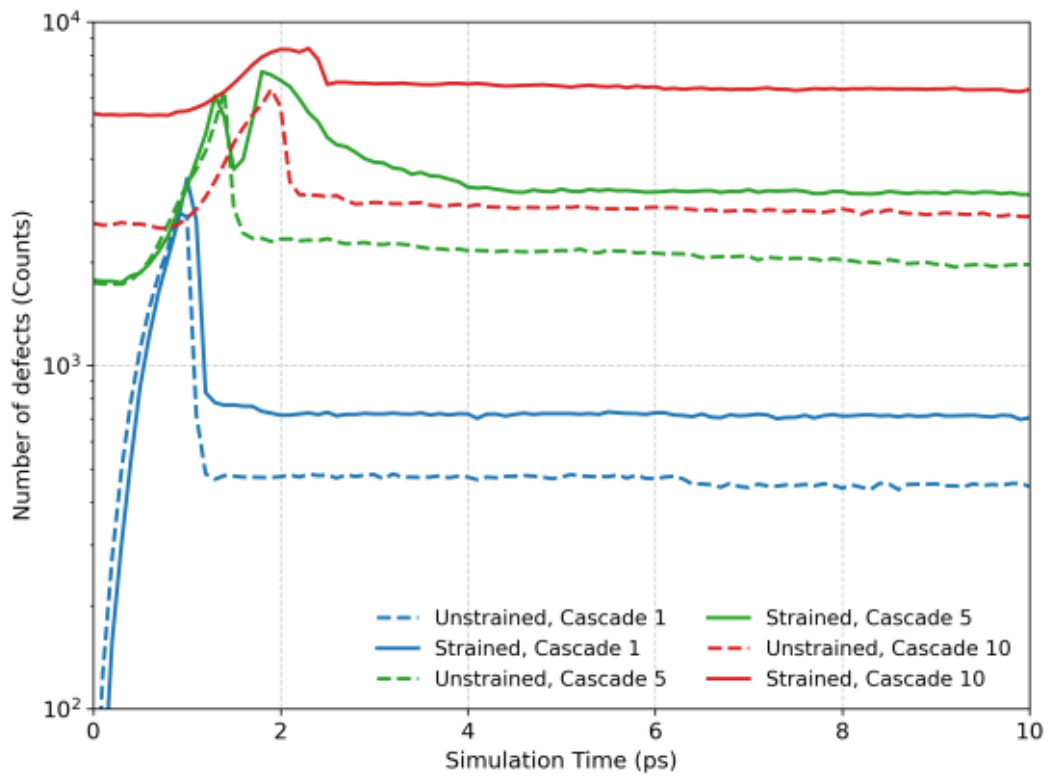

*Fig. 1 Time evolution of defect formation during overlapping 5 keV collision cascades in Ni at 300 K. Unstrained Ni showing the typical rise and decay of defect population associated with the ballistic and recovery phases of individual cascades. Strained Ni illustrating the influence of applied tensile strain, which enhances defect survival and modifies the heat-spike relaxation through stress-assisted defect mobility.*

The unstrained configuration exhibits the typical temporal behavior of a 5 keV collision cascade: a sharp rise in defect population during the ballistic phase, followed by a rapid decay associated with defect recombination during the thermal spike. Using the NRT model, a 5 keV recoil in Ni corresponds to approximately 50 Frenkel pairs (Nordlund et al. 2018; Nair et al. 2025), consistent with the 54 defects obtained directly from our simulations. The cascade 1 was performed after applying a nominal engineering strain of 0.02 (i.e., 2%) along the tensile direction. This strain value lies within the elastic regime of nickel single crystals at 300 K, ensuring that no permanent plastic deformation or dislocation activity occurs prior to irradiation (**Cichocki_Dominguez-**

**Gutierrez_Wyszkowska_Kurpaska_Muszka_2025?**). Subsequent cascades were carried out after applying additional incremental tensile strains. In contrast, the strained systems display noticeable deviations from this trend. The applied strain modifies the local atomic potential energy landscape and enhances defect survival by influencing the recombination dynamics and altering the dissipation of the heat spike. This effect is attributed to stress-assisted defect mobility, where tensile stress promotes preferential migration of interstitials and vacancies along specific crystallographic directions, reducing the likelihood of mutual annihilation. As a result, the defect population tends to stabilize at higher values with increasing strain. Nevertheless, the recovery phase is consistently well reproduced across all cascades, indicating that the model effectively captures both the transient and steady-state aspects of defect evolution under simultaneous mechanical and irradiation conditions. The influence of the applied strain on defect evolution arises from the elastic interaction between point defects and the imposed deformation field. Both vacancies and self– interstitial atoms possess characteristic formation volumes and elastic dipole tensors, so their formation and migration energetics are modified by the external stress state. Under tensile strain, interstitial configurations that locally expand the lattice become energetically more favorable, while vacancy–related configurations become comparatively less favored (Aligayev et al. 2025). This elastic coupling alters the recombination dynamics during the heat spike and increases the survival probability of certain defect species, thereby influencing the overall defect production and dislocation evolution during each cascade.

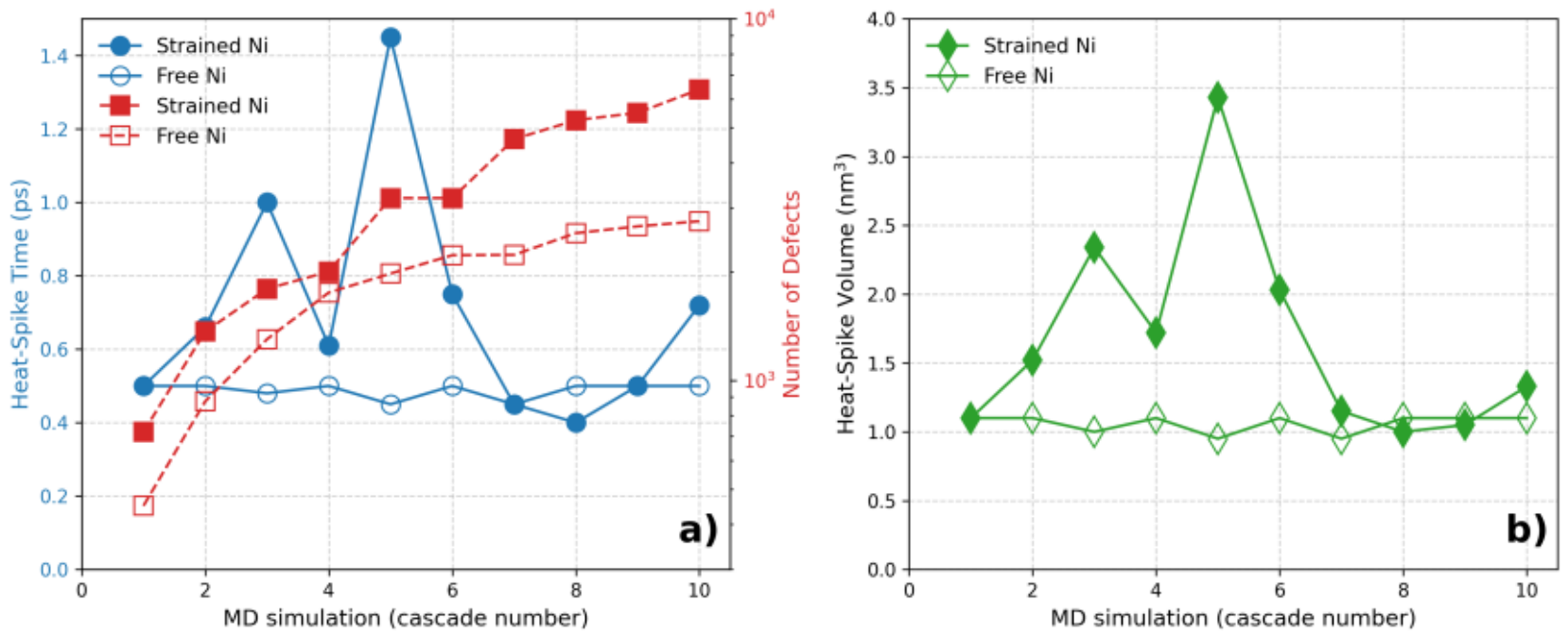

*Fig 2. Heat spike duration and total defect production in a) and Volume in b) as a function of applied tensile strain in Ni.*

Figure 2a) presents the computed heat spike duration and the corresponding total number of defects produced at the end of each collision cascade as a function of the applied tensile strain. The results reveal a clear coupling between defect generation and the thermal response of the system. For the unstrained Ni case, the heat spike duration remains nearly constant across successive overlapping cascades, indicating that the thermal behavior is primarily governed by the energy of the recoil events rather than by cumulative defect production. The lower defect density observed in the unstrained configuration arises from more efficient recombination between self-interstitial atoms and vacancies during the recovery phase. In contrast, the application of tensile strain modifies the local potential energy landscape, introducing internal stress fields that hinder perfect recombination, resulting in enhanced defect retention and a strain-dependent reduction in heat spike duration. As strain increases, both the intensity and lifetime of the heat spike systematically decrease. This reduction reflects the influence of pre-existing strain fields on the local energy dissipation mechanisms during the cascade. The presence of strain promotes stress-assisted defect mobility, allowing interstitials and vacancies to migrate more efficiently along energetically favorable directions. This enhanced mobility facilitates faster relaxation of the transiently disordered region, thereby shortening the duration of the heat spike. Consequently, the molten-like zone formed during the ballistic phase becomes smaller and less stable, leading to a modest decrease in the total number of surviving defects. The applied tensile strain significantly alters the dynamics of irradiation-induced defects by introducing directional stress fields that bias defect mobility. While stress-assisted defect mobility might intuitively suggest enhanced recombination, our simulations indicate the opposite effect—an increase in the surviving defect population. Under tensile stress, the local potential energy landscape becomes asymmetric: interstitial-type defects experience reduced attractive interactions with vacancies, and their migration paths become directionally biased along tensile axes. This anisotropy lowers the probability of mutual annihilation between self-interstitial atoms (SIAs) and vacancies, even though their individual mobilities increase. Moreover, tensile strain stabilizes open-volume regions that energetically favor vacancy retention, while promoting the glide and rearrangement of interstitials into extended structures such as prismatic dislocation loops. As a result, the overall defect population remains higher than in the unstrained case, reflecting a competition between enhanced mobility and strain-induced spatial separation of defects. These findings highlight the critical role of mechanical stress in modulating the defect production and thermal evolution during irradiation events. In Fig. 2b), the effective heat- spike volume is shown and estimated from the peak number of non-fcc atoms during each cascade, converted using the atomic volume of Ni. For the first unstrained 5 keV cascade the spike volume is $\sim 1.1\ \mathrm{nm}^3$. In the unstrained overlap sequence, the heat-spike lifetime is invariant across ten cascades, implying reproducible cascade morphology and therefore an essentially constant spike volume from cascade to cascade. In contrast, the strained sequence exhibits marked cascade-to-cascade variability: the spike lifetime increases by $\sim 2\times$ at cascade 3 and $\sim 3\times$ at cascade 5, while cascade 4 relaxes similarly to cascade 1. This indicates a history-dependent cascade response under tensile strain, where overlap with previously strain-stabilized interstitial-rich clusters yields transiently larger and longer-lived disordered zones. Consequently, the strained system shows intermittent increases in spike volume that track the lifetime enhancements, whereas the unstrained case maintains a nearly constant spike morphology throughout the sequence.

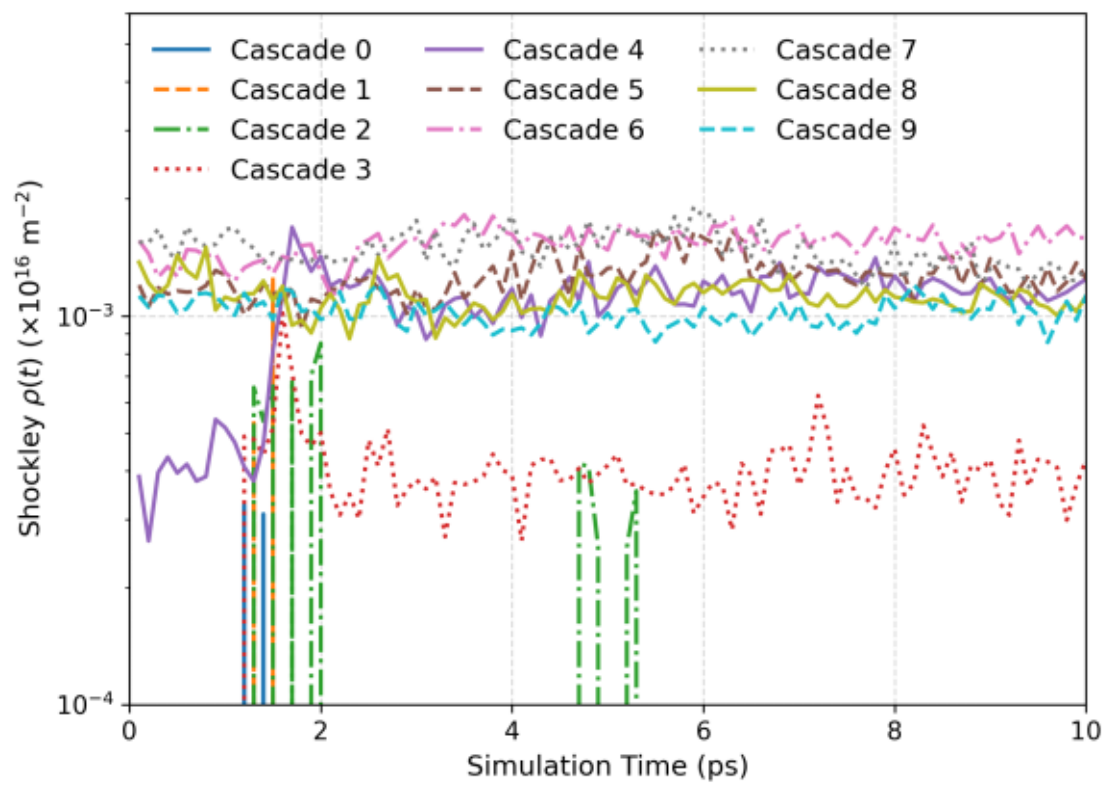

*Fig 3. Evolution of Shockley dislocation density during overlapping collision cascades in unstrained Ni at 300 K.*

Figure 3 presents the evolution of the total and partial dislocation densities, including Shockley, Hirth, Stair-rod, Frank, and Perfect types, for the case of overlapping collision cascades in unstrained nickel. Among the different types, the formation of Frank and Perfect dislocations is negligible, indicating that these configurations are energetically unfavorable under the current irradiation conditions. In contrast, Shockley-type dislocations dominate the overall response, followed by a smaller yet noticeable contribution from Hirth and Stair-rod types. The predominance of Shockley partials, with densities on the order of $10^{-4}$, suggests that the primary defect structures generated in the absence of external strain are prismatic dislocation loops. This interpretation is consistent with previously reported atomistic and experimental studies, where Shockley and Hirth partial dislocations have been identified as key components of loop formation in fcc metals under cascade-induced damage. The relatively low density of other dislocation types supports the conclusion that the microstructural evolution is governed by the nucleation and stabilization of Shockley-type prismatic loops as the dominant recovery mechanism following overlapping cascades.

Figure 4 shows the evolution of the dislocation density for the different types – Shockley, Hirth, Stair-rod, Frank, and Perfect – as a function of simulation time for strained Ni under overlapping collision cascades. The results demonstrate that the applied tensile strain markedly enhances the formation of Shockley dislocations compared to the unstrained case (note the difference in $y$-axis scale), while the density of Hirth-type dislocations decreases by approximately one order of magnitude. This reduction can be attributed to the effect of stress-assisted defect mobility: the applied strain promotes preferential glide of partial dislocations and facilitates their rearrangement into more stable Shockley configurations, thereby suppressing the formation of sessile Hirth dislocations. As the applied strain increases, both the total and Shockley dislocation densities approach a saturation level of about $10^{-2}$. This behavior suggests that the material reaches a steady microstructural state in which the rate of dislocation generation and annihilation becomes balanced. The accumulation and stabilization of Shockley partials at this level indicate that under continued strain, prismatic dislocation loops and extended stacking faults dominate the defect microstructure in irradiated Ni.

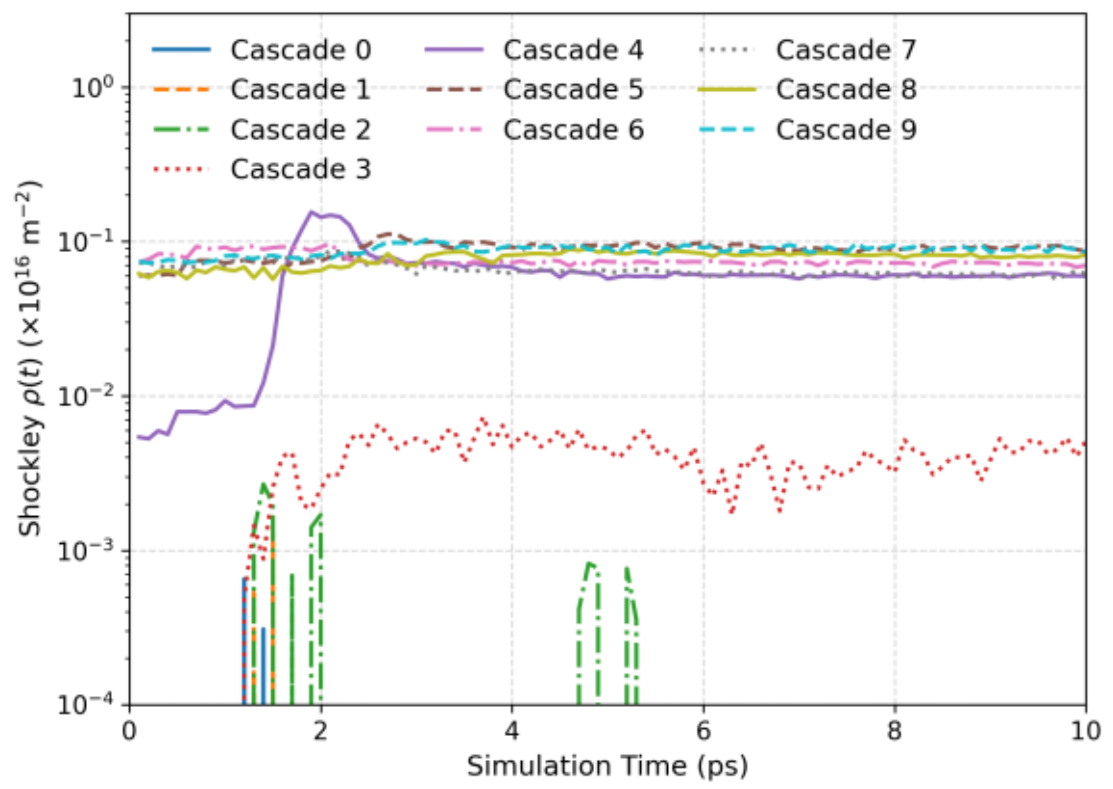

*Fig. 4 Shockley dislocation density evolution as a function of simulation time for strained Ni under overlapping collision cascades.*

To further interpret the evolution of the dislocation density under irradiation and strain, the results were analyzed within the framework of the Kocks–Mecking model, which describes the balance between dislocation storage and dynamic recovery during plastic deformation. In this model, the evolution of the dislocation density ( $\rho$ ) with respect to strain ( $\varepsilon$ ) is given by (Mecking and Kocks 1981)

$$\frac{d\rho}{d\varepsilon} = k_1\sqrt{\rho} - k_2\rho,$$

where ( $k_1$ ) represents the dislocation accumulation coefficient and ( $k_2$ ) accounts for dynamic recovery mechanisms such as annihilation and rearrangement. The steady-state dislocation density predicted by this model is ( $\rho_{ss} = (k_1/k_2)^2$ ), corresponding to the condition where dislocation generation and recovery rates are balanced.

Figure 5 presents the total dislocation density as a function of the applied strain for the case of strained Ni under overlapping collision cascades. A fitting curve based on the Kocks–-Mecking relation was superimposed on the simulation data, yielding a steady-state density of approximately ( $\rho_{ss} = 0.1 \times 10^{16} m - 2$ ). This saturation behavior reflects the stabilization of the microstructure at high strain levels, where the continuous formation and annihilation of dislocations reach equilibrium (Dominguez-Gutierrez et al. 2022). The good agreement between the simulation data and the Kocks–Mecking fit demonstrates that the model can successfully capture the strain-dependent evolution of dislocation structures even under irradiation conditions, emphasizing the strong coupling between mechanical deformation and radiation-induced defect dynamics in nickel. The good agreement between the simulation data and the simplest Kocks–Mecking fit demonstrates that the model can successfully capture the strain-dependent evolution of dislocation structures even under irradiation conditions, emphasizing the strong coupling between mechanical deformation and radiation-induced defect dynamics in nickel. Furthermore, the Kocks-Mecking model can be extended with Arrhenius-type law to incorporate explicit thermal and strain rate effects (Lemoine et al. 2016; Khvan et al. 2025) However, this will introduce additional fitting parameters (e.g., activation energies, strain rate sensitivity) and typically require a systematic exploration of temperature and loading conditions to ensure reliable parameter identification

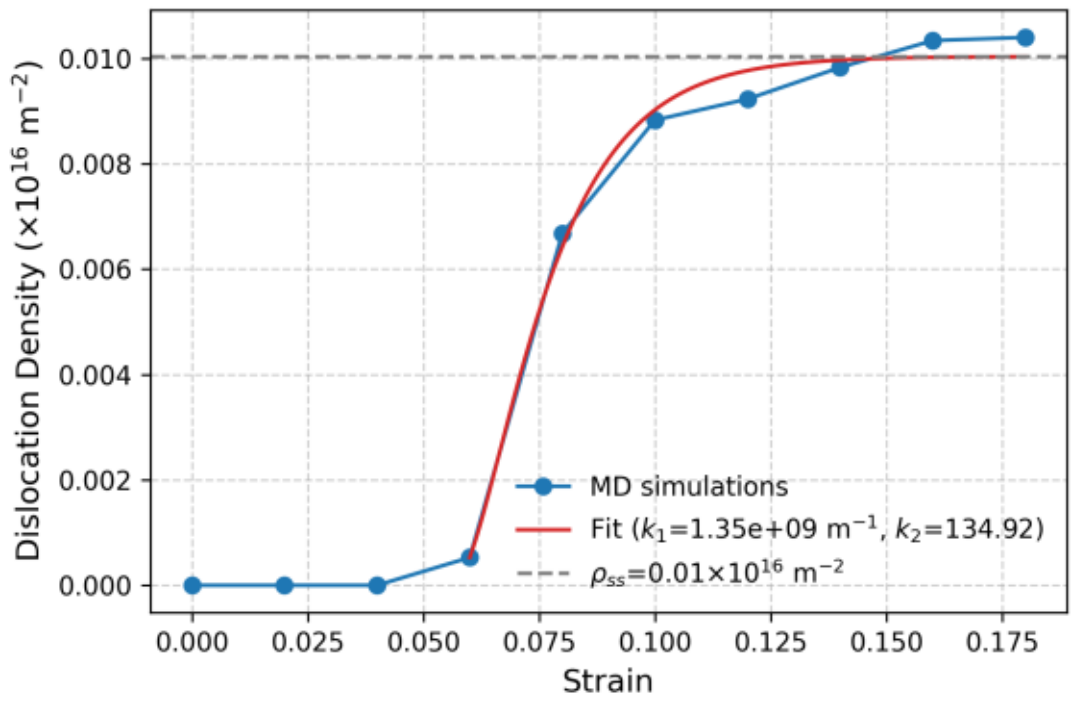

*Fig. 5 Total dislocation density as a function of applied strain, fitted using the Kocks–Mecking model.*

Although the present simulations are performed at a PKA energy of 5 keV, the mechanisms identified in this work are not specific to this energy scale. At higher PKA energies, displacement cascades in fcc metals tend to partition into multiple subcascades, each of which locally resembles a few-keV cascade in terms of temperature, spatial extent, and relaxation time. The strain-dependent defect survival observed here arises from the elastic coupling between point defects and the applied stress field, which is governed by the defect formation volumes and elastic dipole tensors rather than by the total cascade energy (Ustrzycka et al. 2024, 2025; Nair et al. 2025; Crocombette et al. 2016). Similarly, the dominance of Shockley partial dislocations is controlled by the fcc crystallography and stacking-fault energy of Ni, and is expected to persist at higher energies, even though the absolute loop densities and spatial distributions may change. The observed saturation behaviour of the dislocation density reflects a generic balance between defect accumulation and dynamic recovery, consistent with the Kocks–Mecking framework. Increasing the cascade energy would primarily affect the rate at which this steady state is approached, without altering the qualitative existence of a strain-dependent saturation regime.

# 4 Discussion

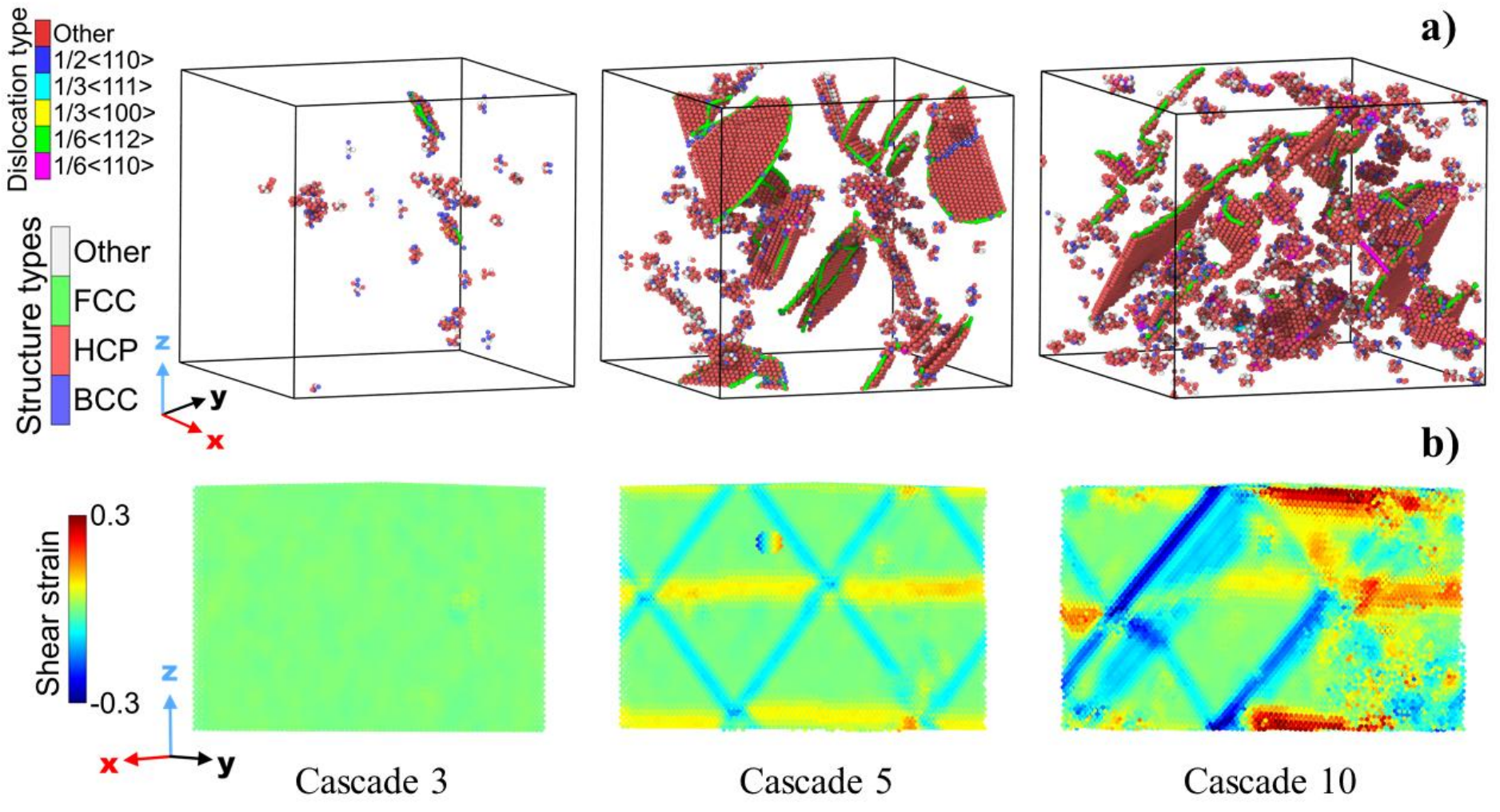

*Fig 6. Atomic snapshots of strained overlap cascades 3, 5, and 10 with corresponding shear–strain mapping, showing strain-assisted defect clustering and the preferential nucleation of Shockley-partial dislocation loops.*

Figure 6 illustrates the atomic structure and dislocations nucleated for the strained overlap cascades (cascades 3, 5, and 10) in a) together with the corresponding atomic shear–-strain maps in b) where the samples are sliced on the [110] plane, providing a direct view of how the deformation field interacts with irradiation damage. Under tensile loading, the local shear strains that develop within and around the cascade core align preferentially along the fcc ($\{111\}\langle 110\rangle$) slip geometry, which is precisely the lattice setting where irradiation–produced interstitial clusters transform into faulted prismatic loops. Consequently, the strained cascades show an enhanced tendency for Shockley–partial dislocation loops to nucleate and grow, while Hirth and stair–rod segments appear mainly as secondary junction products formed by partial–-partial interactions (Ustrzycka et al. 2024, 2025), and Frank loops can be nucleated after a series of cascades and applied strain, as shown in cascade 10. The applied stress not only biases point defect survival toward interstitial rich configurations, but also increases the probability that these clusters shear into extended stacking fault ribbons bounded by Shockley partials; once formed, these loops interact elastically with subsequent cascades, locally concentrating shear and promoting further partial loop accumulation. In short, Fig. 6 links the strain-assisted defect retention seen in the kinetics to a concrete microstructural pathway: tensile-biased cascade disorder produces shear-localized regions that preferentially stabilize Shockley-dominated loop populations and their associated junction products, governing the evolving dislocation network under simultaneous irradiation and deformation.

In addition, the total irradiation dose associated with the ten–cascade sequence used in this work can be estimated using the NRT model. A 5 keV recoil in Ni produces 54 defects in average obtained directly from our MD simulations. Given the 275520 atoms in the simulation cell, this corresponds to a total dose of roughly 0.002 dpa (Ustrzycka et al. 2024). This places the present study firmly within the low–dose regime, where defect production and early–stage recovery dominate and long–range cascade overlap is negligible. It is important to note that the same nominal dpa applies to both strained and unstrained simulations; however, the effective defect accumulation differs markedly between the two cases. In the strained system, the externally imposed deformation field lowers the formation energy of interstitials and enhances their survival during the heat–spike recovery, while suppressing vacancy retention. As a result, the strained material exhibits a higher retained defect fraction and a more pronounced buildup of Shockley–partial dislocations at the same formal dpa. These observations highlight that mechanical loading can significantly bias the primary–damage state even at extremely low doses, and that the qualitative trends reported here reflect strain–assisted modifications to cascade physics rather than differences in irradiation dose.

Furthermore, the present simulations focus on a [100]- oriented Ni single crystal, but the mechanisms identified here are expected to generalize to other crystallographic orientations. In fcc metals, Shockley partial loops are favored because of the low stacking–fault energy and the dissociation of perfect dislocations into partials on {111} planes. Changing the loading direction—for example, to [111] or [110]–would modify the resolved shear stresses acting on the {111} slip planes and consequently alter the relative ease of forming Shockley versus Hirth loops. Although the quantitative balance between these loop types may therefore depend on orientation, the underlying mechanism of strain-assisted stabilization of interstitial-rich clusters remains the same. In polycrystalline or textured materials, different grains will experience different local orientations with respect to the applied stress, leading to grain-dependent loop populations. Nevertheless, each grain obeys the same orientation-dependent elastic interactions between point defects and the external strain field. As a result, the qualitative trends reported here—enhanced interstitial retention, preferential formation of Shockley partial dislocations, and strain-dependent saturation behaviour—are expected to persist across grains, even though the magnitude of these effects will vary with crystallographic orientation. The present study isolates the interaction between applied strain and primary cascade damage in a single–crystal environment, but structural materials contain efficient sinks for point defects such as grain boundaries, free surfaces, and second–phase particles. These sinks can strongly influence the post-cascade evolution by providing pathways for long-range defect migration and annihilation. In the presence of such sinks, the applied strain would still modify defect formation energies and enhance the survival of interstitial–rich clusters during the heat spike, but the subsequent migration of point defects toward nearby sinks would reduce the steady-state defect population compared to the bulk case. The balance between strain–stabilized loop formation and sink absorption would depend on the sink geometry and crystallographic orientation, producing grain-dependent defect retention in polycrystalline materials. The present framework can be extended to examine this competition by including a simple sink within the MD cell, such as a free surface, a coherent interface, or a symmetric tilt grain boundary. Such simulations would enable direct assessment of how strain biases the partitioning of cascade-generated defects between local clustering within the cascade core and absorption at extended defects. This would provide a clearer bridge between the single-crystal mechanisms identified here and the defect evolution expected in polycrystalline and engineering-grade Ni–based alloys (Osetsky et al. 2015).

Simulations were performed at 300 K in order to isolate the intrinsic interaction between the applied elastic strain and the primary displacement cascade, without the additional complexity of thermally activated recovery mechanisms. Performing cascade simulations at elevated temperatures would require substantially greater computational time, as long–range defect migration and annealing processes unfold on timescales far beyond the ballistic and thermal-spike regimes captured in this work. Furthermore, increasing the temperature significantly complicates defect detection: thermal vibrations obscure the local atomic environments, requiring more sophisticated analysis methods and carefully calibrated detection thresholds to distinguish radiation-induced defects from thermal noise. As discussed in detail in our previous work (Suárez-Recio et al. 2025), robust identification of point defects and small clusters at high temperature demands additional post-processing and often quenched configurations. For these reasons, the present study focuses on the fundamental strain–cascade physics at low temperature, while the extension to elevated-temperature conditions is identified as a direction for future research.

# 5 Conclusion

In this work, molecular dynamics simulations were performed to investigate the interplay between mechanical strain and irradiation-induced defect evolution in nickel single crystals subjected to cumulative overlapping 5 keV collision cascades at 300 K. The results reveal a strong coupling between the applied strain field and the defect production, recovery, and dislocation evolution processes.

The time-dependent defect analysis showed that tensile strain modifies the morphology and lifetime of the heat spike, promoting stress-assisted defect mobility and altering recombination dynamics. As a result, strained configurations exhibit enhanced defect survival and reduced heat spike duration compared to the unstrained case. The dislocation analysis further demonstrated that Shockley-type partials dominate the microstructural response during cascade overlap, while Hirth and other dislocation types remain comparatively scarce. Under increasing strain, both the total and Shockley dislocation densities reach a saturation value on the order of $0.1 \times 10^{16}$ m$^{-2}$, indicating the establishment of a steady-state microstructure characterized by stable prismatic loops and extended stacking faults. Furthermore, the evolution of the total dislocation density with applied strain was successfully described using the Kocks–Mecking model. The fitted parameters revealed a steady-state density consistent with the simulation data, confirming that the balance between dislocation accumulation and dynamic recovery governs the observed saturation behavior. These findings provide atomistic insight into the mechanisms governing defect accumulation and microstructural stabilization in strained metals under irradiation, offering a predictive basis for understanding radiation–mechanical coupling in structural materials relevant to nuclear and extreme environments. Validation of the MD cascade results was established through several independent checks. First, the number of surviving Frenkel pairs per 5 keV cascade obtained from Wigner–Seitz analysis is 54, in close agreement with the NRT prediction of approximately 50 Frenkel pairs using the accepted threshold displacement energy for Ni. Second, the temporal defect profiles demonstrate that the ballistic phase, heat spike, and recovery complete within $\lesssim 10$ ps, and extending the simulations beyond 20 ps does not modify the final defect populations, confirming that cascade relaxation is fully captured. Third, the microstructural response is physically consistent with fcc defect mechanics, showing Shockley-partial loop dominance and SIA-rich clustering expected for Ni. Fourth, statistical robustness was ensured by repeating each ten-cascade overlap sequence ten times with randomly selected PKAs and reporting ensemble averages; the associated standard deviations are much smaller than the systematic strain-induced differences. Finally, the main qualitative trends (interstitial-dominated defect survival and Shockley-loop prevalence, enhanced under tensile strain) are consistent with experimental post-irradiation observations in fcc Ni and Ni-based alloys, supporting the physical realism of the strain–cascade coupling captured by the simulations.

Although the present work focuses on pure Ni, the underlying mechanisms identified here are broadly characteristic of fcc metals subjected to irradiation under an applied elastic strain. The strain-induced modification of point-defect energetics, the asymmetric stabilization of interstitials relative to vacancies, and the enhanced formation of Shockley-type partial loops arise from fundamental features of fcc crystallography—such as the defect formation volumes, elastic dipole tensors, and the low stacking-fault energy that governs dislocation dissociation. These mechanisms are not specific to elemental Ni and are expected to persist, qualitatively, in Ni-based alloys and other fcc systems, even in the presence of solute atoms or microstructural heterogeneities. While alloying additions may influence quantitative defect retention or migration pathways, the strain-assisted biasing of defect survival and the corresponding evolution of irradiation-induced dislocation structures represent generic phenomena relevant to a wide class of fcc materials exposed to concurrent mechanical loading and irradiation.

**Acknowledgements**

Research was funded through the European Union Horizon 2020 research and innovation program under Grant Agreement No. 857470 and the initiative of the Ministry of Science and Higher Education "Support for the activities of Centers of Excellence established in Poland under the Horizon 2020 program" under Agreement No. MEiN/2023/DIR/3795. This research is part of the project No. 2022/47/P/ ST5/01169 co-funded by the National Science Centre and the European Union Framework Programme for Research and Innovation Horizon 2020 under the Marie Skłodowska-Curie grant agreement No. 945339. For the purpose of Open Access, the author has applied a CC-BY public copyright license to any Author Accepted Manuscript (AAM) version arising from this submission. We gratefully acknowledge Polish high-performance computing infrastructure PLGrid (HPC

Center: ACK Cyfronet AGH) for providing computer facilities and support within computational Grant No. PLG/2024/017084.

Aligayev, A, M Landeiro Dos Reis, A Chartier, U Jabbarli, F J Dominguez-Gutierrez, and Q Huang. 2025. “Irradiation-Induced Defect Evolution in Concentrated Solid-Solution Alloys: A Molecular Dynamics Perspective.” *Plasma Physics and Controlled Fusion* 67 (5): 055020. https://doi.org/10.1088/1361-6587/adcd2b.

Azeem, M. Mustafa, Zhongyu Li, Qingyu Wang, and Muhammad Zubair. 2018. “Molecular Dynamics Studies and Irradiation Effects in ODSS Alloys.” *International Journal of Nuclear Energy Science and Technology (IJNEST)* 12 (4).

Bonny, G, D Terentyev, R C Pasianot, S PoncĂ©, and A Bakaev. 2011. “Interatomic Potential to Study Plasticity in Stainless Steels: The FeNiCr Model Alloy.” *Modelling and Simulation in Materials Science and Engineering* 19 (8): 085008. https://doi.org/10.1088/0965-0393/19/8/085008.

Chaouadi, R. 2008. “Effect of Irradiation-Induced Plastic Flow Localization on Ductile Crack Resistance Behavior of a 9%cr Tempered Martensitic Steel.” *Journal of Nuclear Materials* 372 (2–3): 379–90.

Crocombette, Jean-Paul, Laurent Van Brutzel, David Simeone, and Laurence Luneville. 2016. “Molecular Dynamics Simulations of High Energy Cascade in Ordered Alloys: Defect Production and Subcascade Division.” *Journal of Nuclear Materials* 474: 134–42. https://doi.org/https://doi.org/10.1016/j.jnucmat.2016.03.020.

Dethloff, Christian, Ermile Gaganidze, and Jarir Aktaa. 2016. “Microstructural Defects in EUROFER97 After Different Neutron Irradiation Conditions.” *Nuclear Materials and Energy* 9: 471–75.

Dominguez-Gutierrez, F J, A Ustrzycka, Q Q Xu, R Alvarez-Donado, S Papanikolaou, and M J Alava. 2022. “Dislocation Nucleation Mechanisms During Nanoindentation of Concentrated FeNiCr Alloys: Unveiling the Effects of Cr Through Molecular Simulations.” *Modelling and Simulation in Materials Science and Engineering* 30 (8): 085010. https://doi.org/10.1088/1361-651X/ac9d54.

Fazio, Concetta, Dolores Gomez Briceno, Michael Rieth, Alessandro Gessi, Jean Henry, and Lorenzo Malerba. 2011. “Innovative Materials for Gen IV Systems and Transmutation Facilities: The Cross-Cutting Research Project GETMAT.” *Nuclear Engineering and Design* 241 (9): 3514–20.

Gaganidze, E., and J. Aktaa. 2013. “Assessment of Neutron Irradiation Effects on RAFM Steels.” *Fusion Engineering and Design* 88 (3): 118–28.

Gao, F., and W. J. Weber. 2003. “Atomic Modeling of Defects, Defect Generation and Multiple Ion-Solid Interactions.” *AIP Conference Proceedings* 680: 575–78. https://doi.org/10.1063/1.1619799.

Garner, F., M. Hamilton, N. Panayotou, and G. Johnson. 1981. “The Microstructural Origins of Yield Strength Changes in Aisi 316 During Fission or Fusion Irradiation.” *Journal of Nuclear Materials* 104: 803–7.

Gentils, A., and C. Cabet. 2019. “Investigating Radiation Damage in Nuclear Energy Materials Using JANNuS Multiple Ion Beams.” *Nuclear Instruments and Methods in Physics Research Section B: Beam Interactions with Materials and Atoms* 447 (May): 107–12. https://doi.org/10.1016/j.nimb.2019.02.022.

Guenole, Julien, Wolfram G. Noehring, Aviral Vaid, et al. 2020. “Assessment and Optimization of the Fast Inertial Relaxation Engine (Fire) for Energy Minimization in Atomistic Simulations and Its Implementation in Lammps.” *Computational Materials Science* 175: 109584. https://doi.org/https://doi.org/10.1016/j.commatsci.2020.109584.

Jagielski, Jacek, and Lionel Thomé. 2009. “Multi-Step Damage Accumulation in Irradiated Crystals.” *Applied Physics A* 97: 147–55.

Khater, H., G. Monnet, D. Terentyev, and A. Serra. 2014. “Dislocation Glide in Fe–Carbon Solid Solution: From Atomistic to Continuum Level Description.” *International Journal of Plasticity* 62: 34–49.

Khvan, Tymofii, Ludovic Noels, Dmitry Terentyev, Enrico Corniani, Peter HĂ¤hner, and Chih-Cheng Chang. 2025. “Development of a Semi-Empirical Approach Correlating Ion- and Neutron-Induced Hardening in Eurofer97 Using Nanoindentation and Crystal Plasticity Finite Element Method.” *Journal of Materials Research and Technology* 38: 5780–96. https://doi.org/https://doi.org/10.1016/j.jmrt.2025.08.235.

Kohyama, A., Y. Kohno, K. Asakura, M. Yoshino, C. Namba, and C. Eiholzer. 1994. “Irradiation Creep of Low-Activation Ferritic Steels in FFTF/MOTA.” *Journal of Nuclear Materials* 212–215: 751–54.

Kumar, N. Naveen, P. V. Durgaprasad, B. K. Dutta, and G. K. Dey. 2012. “Modeling of Radiation Hardening in Ferritic/Martensitic Steel Using Multi-Scale Approach.” *Computational Materials Science* 53 (1): 258–67.

Kurpaska, L., F. J. Dominguez-Gutierrez, Y. Zhang, et al. 2022. “Effects of Fe Atoms on Hardening of a Nickel Matrix: Nanoindentation Experiments and Atom-Scale Numerical Modeling.” *Materials & Design* 217: 110639. https://doi.org/https://doi.org/10.1016/j.matdes.2022.110639.

Lemoine, Guerric, Laurent Delannay, Hosni Idrissi, Marie-StĂ©phane Colla, and Thomas Pardoen. 2016. “Dislocation and Back Stress Dominated Viscoplasticity in Freestanding Sub-Micron Pd Films.” *Acta Materialia* 111: 10–21. https://doi.org/https://doi.org/10.1016/j.actamat.2016.03.038.

Levo, E., F. Granberg, C. Fridlund, K. Nordlund, and F. Djurabekova. 2017. “Radiation Damage Buildup and Dislocation Evolution in Ni and Equiatomic Multicomponent Ni-Based Alloys." *Journal of Nuclear Materials* 490: 323–32. https://doi.org/https://doi.org/10.1016/j.jnucmat.2017.04.023.

Lin, Pandong, Junfeng Nie, and Meidan Liu. 2022. "Multiscale Crystal Plasticity Finite Element Model for Investigating the Irradiation Hardening and Defect Evolution Mechanism of A508-3 Steel." *Nuclear Materials and Energy* 32: 101214.

Malerba, L., M. Caturla, E. Gaganidze, et al. 2021. "Multiscale Modelling for Fusion and Fission Materials: The M4F Project." *Nuclear Materials and Energy* 29.

Massoud, Jean-Paul, Stéphane Bugat, Bernard Marini, David Lidbury, and Steven Van Dyck. 2010. "PERFECT – Prediction of Irradiation Damage Effects on Reactor Components: A Summary." *Journal of Nuclear Materials* 406 (1): 2–6. https://doi.org/10.1016/j.jnucmat.2010.06.009.

Mecking, H., and U. Kocks. 1981. "Kinetics of Flow and Strain-Hardening." *Acta Metallurgica* 29: 1865–75.

Mieszczynski, C., E. Wyszkowska, P. Jozwik, et al. 2024. "Damage Kinetics in High-Temperature Irradiated Ni Crystals." *Applied Surface Science* 676: 160991. https://doi.org/https://doi.org/10.1016/j.apsusc.2024.160991.

Mieszczynski, Cyprian, Kazimierz Skrobas, Przemyslaw Jozwik, et al. 2025. "Dislocation Loops Parameterization for Ion Channeling Analysis of High-Temperature Ar-Implanted NiFe and NiFeCoCr Alloys." *Physica Status Solidi (RRL) " Rapid Research Letters* n/a (n/a): 2500083.

Nair, Adithya, Charlotte S. Becquart, Christophe Domain, and AndrÃ©e De Backer. 2025. "Statistical Study of Displacement Cascades in Ni and FeNiCr Alloys: Understanding the Influence of Potential and Composition on Primary Damage Modeling." *Journal of Nuclear Materials* 612: 155832. https://doi.org/https://doi.org/10.1016/j.jnucmat.2025.155832.

Nordlund, Kai, Steven J. Zinkle, Andrea E. Sand, et al. 2018. "Primary radiation damage: A review of current understanding and models." *Journal of Nuclear Materials* 512: 450–79. https://doi.org/{https://doi.org/10.1016/j.jnucmat.2018.10.027}.

Odette, G., and G. Lucas. 1997. "Recent Progress in Understanding Reactor Pressure Vessel Steel Embrittlement." *Radiation Effects and Defects in Solids* 144 (1–4): 189–231.

Osetsky, Y., N. Anento, A. Serra, and D. Terentyev. 2015. "The Role of Nickel in Radiation Damage of Ferritic Alloys." *Acta Materialia* 84: 368–74. https://doi.org/https://doi.org/10.1016/j.actamat.2014.10.060.

Schäublin, R., Z. Yao, N. Baluc, and M. Victoria. 2005. "Irradiation-Induced Stacking Fault Tetrahedra in Fcc Metals." *Philosophical Magazine* 85 (4–7): 769–77.

Slugeň, V., S. Sojak, W. Egger, V. Krsjak, J. Veternikova, and M. Petriska. 2020. “Radiation Damage of Reactor Pressure Vessel Steels Studied by Positron Annihilation Spectroscopy—a Review.” *Metals* 10.

Stukowski, A. 2009. “Visualization and Analysis of Atomistic Simulation Data with OVITO–the Open Visualization Tool.” *Modelling and Simulation in Materials Science and Engineering* 18 (1): 015012.

Suárez-Recio, J., P. M. Piaggi, F. J. Domínguez-Gutiérrez, et al. 2025. “Early Stages of Self-Healing at Tungsten Grain Boundaries from *Ab Initio* Machine Learning Simulations.” *Communications Materials* 6 (1): 236. https://doi.org/10.1038/s43246-025-00945-6.

Thompson, Aidan P., H. Metin Aktulga, Richard Berger, et al. 2022. “LAMMPS - a Flexible Simulation Tool for Particle-Based Materials Modeling at the Atomic, Meso, and Continuum Scales.” *Computer Physics Communications* 271: 108171. https://doi.org/https://doi.org/10.1016/j.cpc.2021.108171.

Ustrzycka, A., F. J. Dominguez-Gutierrez, and W. Chrominski. 2024. “Atomistic Analysis of the Mechanisms Underlying Irradiation-Hardening in Fe-Ni-Cr Alloys.” *International Journal of Plasticity* 182: 104118. https://doi.org/https://doi.org/10.1016/j.ijplas.2024.104118.

Ustrzycka, A., H. Mousavi, F. J. Dominguez-Gutierrez, and S. Stupkiewicz. 2025. “Atomistic Study of Radiation-Induced Ductile-to-Brittle Transition in Austenitic Steel.” *International Journal of Mechanical Sciences* 303: 110567. https://doi.org/https://doi.org/10.1016/j.ijmecsci.2025.110567.

Wakai, E., S. Takaya, Y. Matsui, et al. 2021. “Irradiation Damages of Structural Materials Under Different Irradiation Environments.” *Journal of Nuclear Materials* 543.

Was, G. 2015. “Challenges to the Use of Ion Irradiation for Emulating Reactor Irradiation.” *Journal of Materials Research* 30: 1158–82.

Won-Mi, Choi, Jo Yong Hee, Sohn Seok Su, Lee Sunghak, and Lee Byeong-Joo. 2018. “Understanding the Physical Metallurgy of the CoCrFeMnNi High-Entropy Alloy: An Atomistic Simulation Study.” *Npj Comput Mater* 4.

Wyszkowska, Edyta, Cyprian Mieszczynski, Amil Aligayev, et al. 2025. “Nanoscale Defect Formation in Fcc Ni and Low-Fe Content NixFe1-x Single Crystals Induced by Self-Ion Irradiation.” *Nanoscale* 17: 15841–55.